\documentclass[journal]{IEEEtran}
\PassOptionsToPackage{ruled, linesnumbered}{algorithm2e}
\usepackage[latin9]{inputenc}
\usepackage{color}
\usepackage{algorithm2e}
\usepackage{amsmath}
\usepackage{amsthm}
\usepackage{amssymb}
\usepackage{graphicx}
\usepackage{cite}
\usepackage{setspace,subcaption}
\usepackage[font=small]{caption}
\allowdisplaybreaks 
\makeatletter
\theoremstyle{plain}
\newtheorem{thm}{\protect\theoremname}
\newtheorem{remark}{Remark}
\newtheorem{lema}{Lemma}

\usepackage{balance}
\usepackage{pgfplots}
\usepackage{pgfplotstable}
\usepackage{tikz}
\usetikzlibrary{calc}
\usetikzlibrary{arrows.meta}
\usetikzlibrary{patterns}

\usepackage{comment}

\makeatother

\providecommand{\theoremname}{Theorem}

\begin{document}
\title{\LARGE{A Robust Design for BackCom Assisted Hybrid NOMA}}
\author{Muhammad Fainan Hanif, Le-Nam Tran,~\IEEEmembership{Senior Member,~IEEE,} Zhiguo Ding,~\IEEEmembership{Fellow,~IEEE,}\\ and Tharmalingam Ratnarajah,~\IEEEmembership{Senior Member,~IEEE}
\thanks{M. F. Hanif is with the Institute of Electrical, Electronics and Computer
Engineering, University of the Punjab, Lahore, Pakistan. (email: fainan.hanif@gmail.com).}	
\thanks{ L.-N. Tran is with the School of Electrical and
Electronic Engineering, University College Dublin, D04 V1W8, Ireland (e-mail: nam.tran@ucd.ie).}
\thanks{ Z. Ding is with the University of Manchester, Manchester, M1 9BB, UK,
and Khalifa University, Abu Dhabi, UAE
(email: zhiguo.ding@manchester.ac.uk).}
\thanks{ T. Ratnarajah is with the Department of Electrical and Computer Engineering (ECE), College of Engineering, San Diego State University, San Diego, CA 92182 USA (email: t.ratnarajah@ieee.org).
}}
\maketitle
\begin{abstract}
Hybrid non-orthogonal multiple access (H-NOMA) is inherently an enabler of massive machine type communications, a key use case for sixth-generation (6G) systems. Together with backscatter communication (BackCom), it seamlessly integrates with the traditional orthogonal multiple access (OMA) techniques to yield superior performance gains. In this paper, we study BackCom assisted H-NOMA uplink transmission with the aim of minimizing power with imperfect channel state information (CSI), where a generalized representation for channel estimation error models is used. The considered power minimization problem with aggregate data constraints is both non-convex and intractable. For the considered imperfect CSI models, we use Lagrange duality and the majorization-minimization principle to produce a conservative approximation of the original problem. The conservative formulation is relaxed by incorporating slack variables and a penalized objective. We solve the penalized tractable approximation using a provably convergent algorithm with polynomial complexity. Our results highlight that, despite being conservative, the proposed solution results in a similar power consumption as for the nominal power minimization problem without channel uncertainties. Additionally, robust H-NOMA is shown to almost always yield more power efficiency than the OMA case. Moreover, the robustness of the proposed solution is manifested by a high probability of feasibility of the robust design compared to the OMA and the nominal one. 
\end{abstract}


\section{Introduction}
Multiple access techniques form the cornerstone of research activities in wireless networks beyond fifth-generation (B5G) and sixth-generation (6G). Among several contenders for standardization, non-orthogonal multiple access (NOMA) is a promising paradigm \cite{DingNGMA}. Recently, under the assumption of perfect channel estimation, NOMA has been investigated from the point of view of integration with traditional orthogonal multiple access (OMA) systems, thus leading to the so-called hybrid NOMA (H-NOMA) technique \cite{DingHNoma,DingMECHNoma}.\par The prominent feature of H-NOMA is its ability to integrate with existing OMA schemes, such as time division multiple access (TDMA). In particular, a user in the H-NOMA setup can communicate in time slots which would be solely occupied by other users using traditional NOMA, and also during its own dedicated time slot using OMA. The main distinguishing feature of H-NOMA is that a user partially offloads its data into the time slots of other users that transmit prior to it and completes its data transmission by the time its dedicated time slot finishes. The foundation of H-NOMA was laid in earlier works, where NOMA was used to facilitate mobile edge computing (MEC) in the uplink \cite{Ding_MEC, Ding-MEC1, Liu-HNoma}. Ding \emph{et al.} have extensively explored H-NOMA transmission in the downlink in \cite{DingDLNoma}. In their work \cite{Cirine}, Chaieb {\it et al.} used deep reinforcement learning for resource allocation for H-NOMA enabled millimeter-wave communications in a multi-band setting. The pairing problem was studied in H-NOMA downlink by Lei {\it et al.} in \cite{Lei} where they optimize the date rate in the presence of simultaneous transmitting and reflecting reconfigurable intelligent surface (STAR-RIS). When dynamic switching between OMA and NOMA is allowed in the downlink, Wang {\it et al.} considered the minimization of the age of information in their research work in \cite{WangAoI}. An alternate optimization algorithm when OMA and NOMA are coupled together in an integrated sensing and communication (ISAC) system, supported by RIS was presented in Lyu {\it et al.} \cite{Lyu}. Zhu {\it et al.} in \cite{ZhuURLLC} have thoroughly explored the key performance metric of ultra-reliable and low-latency communication in the finite blocklength regime by maximizing the effective capacity when NOMA operates together with time-division multiple access (TDMA). Sultana and Dumitrescu \cite{Sultana} proposed a globally optimal solution to a downlink H-NOMA max-min data rate optimization problem by performing joint power and channel allocation such that the users are segregated in clusters. In a recent study, Ding \emph{et al.} \cite{DingHNoma} have proposed H-NOMA in the uplink with users capable of scattering ambient signals using backscatter communication (BackCom) circuits. \par
\emph{Motivation:} A major use case for B5G networks is super massive machine type communication that is envisioned to provide massive connection density \cite{Liebhart}. H-NOMA provides a lucrative opportunity to meet this goal since H-NOMA is inherently capable of allowing multiple users to share a resource block by integrating with the existing OMA architecture. In addition to data rates, the primary source of concern for the huge wireless Internet of Things (IoT) is the capacity of nodes to communicate with minimum power requirements. Wireless sensor nodes operate in environments where power is in short supply, and hence reflected or backscatterd signals \cite{Ping} can be a precious source for powering such devices. Therefore, a combination of H-NOMA with BackCom makes it a viable enabling technqiue for such use cases in future. \par The results of H-NOMA for uplink communication with users with BackCom capabilities \cite{DingHNoma} are based on the fundamental assumption of the availability of perfect channel state information (CSI) at all communication nodes in the network. In real world settings, CSI errors are unavoidable, and hence those schemes relying on perfect CSI are no longer applicable. In addition, the transmission of the BackCom signals results in product of channels in the signal-to-interference plus noise ratio (SINR) terms. Hence, when channel imperfections are incorporated, the resultant SINR expressions are intractable for optimization purposes. Motivated by this, we study the problem of minimizing uplink power transmission when perfect CSI is not fully available to all nodes. In our model of channel imperfections, we consider channel estimation errors to belong to certain regions irrespective of their probabilistic nature. This gives rise to the so-called robust version \cite{AharonNemirovskiGhaoui2009} of the BackCom assisted power minimization problem in H-NOMA settings.\par
\emph{Contributions:} The main contributions of the paper are listed as follows:
\begin{itemize}
\item The BackCom capability of users results in  bilinear product of channels with CSI errors in data rate expressions. Resultantly, the traditional approaches to handle the worst-case robust counterpart of the problem do not work. Hence, to overcome this problem we exploit Lagrange duality to obtain a tractable version of the robust counter part.
\item The performance of H-NOMA is critically dependent on the quality of CSI. Generalizing the works as \cite{Biguesh,FangImp}, we model CSI errors such that they appear as errors in nominal channel gains. Specifically, CSI error terms are collected as a sum of terms that are generalized as shifts in nominal channel gains weighted by perturbation vectors. The perturbation vectors of all channels are assumed to lie in the commonly used uncertainty sets.
\item To make the H-NOMA model more realistic with imperfect CSI, we also incorporate the effect of imperfect successive interference cancellation (SIC). The residual interference due to imperfect SIC is accounted for in the design of our algorithmic solution to the main problem formulated in the paper.
\item Despite leveraging duality, the robust counterpart is non-convex. We use the minorization-majorization  framework to devise a polynomial time provably convergent algorithm. As a consequence, we have a novel combination of robust bilinear optimization with the minorization-majorization principle.
\item Finally, numerical results are presented to show the superior performance of the proposed algorithm for BackCom enabled H-NOMA systems. Our results demonstrate that the H-NOMA transmission not only consumes less power compared to OMA, but also guarantees that the minimum data requirements of users are met for a large proportion of time even in the presence of imperfect CSI and SIC.
\end{itemize}
The remainder of the paper is organized as follows. The setup of the system model, the CSI error model, and the main problem formulation are presented in Sec. \ref{SystemModel}. In Sec. \ref{PropSol}, the algorithmic solution to the proposed problem is developed and its various characteristic features are investigated. Numerical results are presented in Sec. \ref{Res}. Finally, we conclude the paper in Sec. \ref{Conc}.\par
\emph{Notation:} We use lower-case letters for scalars and upper-case letters are reserved for vectors. $\mathbf{v}^T$ represents transpose of a vector $\mathbf{v}$. The real and imaginary parts of a complex number $c$ are denoted by $\Re(c)$ and $\Im(c)$, respectively. The optimal value of a variable $v$ is denoted by $v^\star$.
\section{System Model and Problem Formulation}\label{SystemModel}
\subsection{BackCom Assisted Hybrid NOMA}

\textcolor{black}{We consider a system with $U$ users communicating
	with a base station (BS) in the uplink. It is assumed that each node in the system is equipped with a single antenna. Additionally, each
	user possesses the BackCom capability, i.e., using information signals of other
	users to transmit its own data with the aid of a BackCom circuit.
	The data transmission of all users is assumed to occur within a single
	coherence time period, which is divided into $U$ separate time slots.
	In the conventional TDMA scheme, each user is allocated a dedicated
	time slot $T_u$ for transmission, where we assume
	that user $u\in\mathcal{U}\triangleq\{1,2,\ldots,U\}$ transmits in
	time slot $T_{u}$. The channel between the BS and user $u$ is denoted
	by $h_{u}$, which remains constant across all $U$ time slots. For
	reasons that shall be explained shortly, we assume that the channel
	gains are ordered in a decreasing order: $|h_{1}|^{2}>|h_{2}|^{2}>\cdots>|h_{U}|^{2}$.}

\begin{figure}
\centering
\includegraphics[width=1\columnwidth]{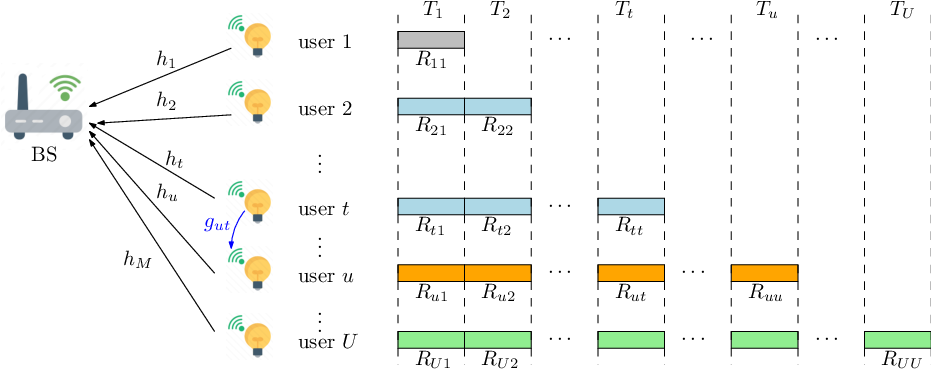}
\caption{Illustration of BackCom assisted hybrid NOMA. During $T_{t}$, users $u>t$ are allowed to transmit their data using the BackCom mode. The data rate of user $u$ during time slot $t\protect\geq u$ is denoted by $R_{ut}$. The channel gains are ordered
in a decreasing order $|h_{1}|^{2}>|h_{2}|^{2}>\cdots>|h_{U}|^{2}$
and SIC is carried out in an increasing order to detect users's signals. BackCom between user $u$ in time slot $t$ and transmission in its dedicated time slot is shown in the same color.} \label{fig:systemmodel}
\end{figure}

\textcolor{black}{H-NOMA has been recently introduced as an additional
	feature of existing TDMA-based systems. Unlike conventional TDMA,
	where only user $u$ transmits in its assigned time slot $T_{u}$,
	H-NOMA allows users $j>u$ to simultaneously transmit their own signals
	in time slot $u$ using BackCom. Note that, the data transmission
	of user $u$ is still completed by the end of $T_{u}$. The considered
	system model is illustrated in Fig. \ref{fig:systemmodel}. }

\textcolor{black}{Consider a specific user $t$. During time slot $T_{t}$,
	user $t$ transmits $x_{tt}$, while users $u>t$ transmit $x_{ut}$
	through BackCom to the BS. Under this setup, the received signal at
	the BS during $T_{t}$ is given by
	\begin{align}		y_{t}=\sqrt{P_{t}}h_{t}x_{tt}+\sum_{u>t}h_{u}g_{ut}x_{tt}\sqrt{\beta_{ut}P_{t}}x_{ut}+n_{t}\label{rxedsig1}
	\end{align}
where $P_{t}$ is the transmit power of
	user $t$ in $T_{t}$, $\beta_{ut}\leq1$ denotes the reflecting coefficient
	of users $u$ in $T_{t}$, $g_{ut}$ is the channel between users $u$
	and $t$, and $n_{t}$ is the additive white Gaussian noise with variance
	$\sigma_{t}^{2}$. }

\textcolor{black}{At the BS, SIC is performed to decode the users'
	signals. Given that the users are ordered in decreasing channel gains,
	it is reasonable to decode the users' signals in an increasing order,
	starting from the first user. By the end of $T_{u-1}$, the signals
	of users $1$ to $u-1$ are successfully decoded. Thus, the data rate
	$\bar{R}_{ut}$ (in b/s/Hz) of user $u$ during time slot $t$, where
	$u\geq t$, is given by 
	\begin{multline}
		\bar{R}_{ut}=\log_{2}\left(1+\frac{|h_{u}|^{2}|g_{ut}|^{2}\beta_{ut}P_{t}}{\sum_{j=u+1}^{U}|h_{j}|^{2}|g_{jt}|^{2}\beta_{jt}P_{t}+\sigma_{t}^{2}}\right),\\
		t=1,2,\ldots,u
	\end{multline}
	where $g_{uu}=\beta_{uu}\triangleq1$. It is now clear that during
	$T_{t}$, user $u$ suffers more interference than any user $j>u$.
	Thus, ordering users in terms of decreasing channel gains makes sequential
	detection in an increasing order more reliable. Without loss of generality,
	we assume that all time slots have the same duration. 
}

\subsection{Channel Error Model}
Perfect CSI at all nodes is not always possible due to factors like mobility, channel estimation errors, etc. Motivated by this, the channels with imperfect CSI are modeled as:{\color{black}
\begin{align}
h_{u} & =\bar{h}_{u}+\Delta h_{u}\label{CSI_err_hu}\\
g_{ut} & =\bar{g}_{ut}+\Delta{g}_{ut}\label{CSI_err_gut}
\end{align}
where $\bar{h}_{u}$ and $\bar{g}_{ut}$ denote the channel estimates,
also called nominal channel values, and $\Delta h_{u},\Delta g_{ut}$
are the channel errors in $h_{u}$ and $g_{ut}$, respectively. \textcolor{black}{The above model leads to the following channel gain errors:
\begin{align}
&|h_u|^2=|\bar{h}_u+\Delta h_u|^2=|\bar{h}_u|^2+2\Re({\bar{h}_u\Delta h_u})+|\Delta h_u|^2\label{actual_channels_or1}\\
&|g_{ut}|^2=|\bar{g}_{ut}+\Delta{g}_{ut}|^2=|\bar{g}_{ut}|^2+2\Re(\bar{g}_{ut}\Delta{g}_{ut})
+|\Delta g_{ut}|^2.\label{actual_channels_or2}
\end{align}
Remarkably, the above channel gains error models can be generalized. The quadratic error terms i.e., $|\Delta h_u|^2$ and $|\Delta g_{ut}|^2$ can be replaced by their maximum values, represented as, $|\Delta h_u|^2_{\max}$ and $|\Delta g_{ut}|^2_{\max}$, respectively. Fixing the quadratic error terms to their maximum values is in line with the concept of the worst-case robust optimization framework \cite{AharonNemirovskiGhaoui2009} that is used in this paper. Therefore, a generalized expression for channel gain errors can be expressed as
\begin{align}
&|h_u|^2=|\bar{h}_u|^2+\sum_{j=1}^L\gamma_j^ua_j^u =|\bar{h}_u|^2+\boldsymbol{\gamma}_u^{T}\boldsymbol{\alpha}_u\label{actual_channels_1}\\
& |g_{ut}|^2=|\bar{g}_{ut}|^2+\sum_{j=1}^L\gamma_j^{ut}b_j^{ut}=
|\bar{g}_{ut}|+\boldsymbol{\gamma}_{ut}^{T}\boldsymbol{\kappa}_{ut}
\label{actual_channels}\end{align}
where $\boldsymbol{\alpha}_u\triangleq[a_1^u,a_2^u,\ldots,a_L^u]^T,\boldsymbol{\kappa}_{ut}\triangleq[b_1^{ut},b_2^{ut},\ldots,b_L^{ut}]^T,\boldsymbol{\gamma}_u\triangleq [\gamma_1^{u},\gamma_2^{u},\ldots,\gamma_L^{u}]$ and $\boldsymbol{\gamma}_{ut}\triangleq [\gamma_1^{ut},\gamma_2^{ut},\ldots,\gamma_L^{ut}]$. To elaborate on this development, if we take $L=3, \boldsymbol{\alpha}_u=[1,2\Re(\bar{h}_u),2\Im(\bar{h_u})]^T$ and $\boldsymbol{\gamma}_u=[|\Delta h_u|^2_{\max},\Re(\Delta h_u),\Im(\Delta h_u)]$, \eqref{actual_channels_1} is reduced to \eqref{actual_channels_or1}. Similarly, \eqref{actual_channels} collapses to \eqref{actual_channels_or2}, when $L=3, \boldsymbol{\kappa}_{ut}=[1,2\Re(\bar{g}_{ut}),2\Im(\bar{g}_{ut})]^T$ and $\boldsymbol{\gamma}_{ut}=[|\Delta g_{ut}|^2_{\max},\Re(\Delta g_{ut}),\Im(\Delta g_{ut})]$. } It is important to note here that the manifestation of CSI errors in channel gains has been considered in earlier works, e.g., \cite{Fang}, \cite{Biguesh}. Interestingly, our model of channel gain errors is much more generalized considering the incorporation of `perturbation vectors' ($\boldsymbol{\gamma_{u}},\boldsymbol{\gamma_{ut}}$) and `shifts' ($\boldsymbol{\alpha}_u,\boldsymbol{\kappa}_{ut}$) shown in the second terms of \eqref{actual_channels_1} and \eqref{actual_channels}, cf. \cite{AharonNemirovskiGhaoui2009}. Our CSI error model is distribution independent. Specifically, the perturbation vectors are assumed to be contained in the uncertainty sets defined as
\begin{align}
&\mathcal{U}^u=\{\boldsymbol{\gamma}_u\in\mathbb{R}^L|\mathbf{A}\boldsymbol{\gamma}_u\geq -\mathbf{l},\mathbf{B}\boldsymbol{\gamma}_u\leq \mathbf{u}\}\label{uncertTyeI}\\& \mathcal{U}^{ut}=\{\boldsymbol{\gamma}_{ut}\in\mathbb{R}^L|\|\boldsymbol{\gamma}_{ut}\|_2\leq \rho\}\label{uncertTyeII}
\end{align}
where we assume that both $\mathbf{A}$ and $\mathbf{B}$ are positive definite matrices, and $\mathbf{l}$ and $\mathbf{u}$ are strictly positive. Under this assumption, the matrices $\mathbf{A}$ and $\mathbf{B}$ can be considered diagonal. Indeed, let $\mathbf{D}_A$ and $\mathbf{D}_B$ be diagonal matrices with eigenvalues of $\mathbf{A}$ and $\mathbf{B}$ on their diagonals, respectively. Noting that $\mathbf{A}^{-1}\mathbf{D}_A$ and $\mathbf{B}^{-1}\mathbf{D}_B$ are both invertible, therefore, their columns form the basis of $\mathbb{R}^L$. Hence, any uncertainty vector given in $\mathbb{R}^L$ belongs to the column space of these matrices. \textcolor{black}{Mathematical tractability is the main motivation of choosing the uncertainty sets in \eqref{uncertTyeI} and \eqref{uncertTyeII}. In particular, unlike the traditional approach, \eqref{uncertTyeI} has been chosen as a polyhedral set instead of a spherical one. We emphasize that \eqref{uncertTyeI} and \eqref{uncertTyeII} are quite generic in their modeling capabilities. In fact, if the parameters in the set \eqref{uncertTyeI} are chosen properly, this uncertainty set can be used to closely approximate any convex body, such as a sphere or an ellipsoid \cite{Marton,Bronshteyn}. The versatility of \eqref{uncertTyeI} can be observed by noting the fact that when $\mathbf{A}=\mathbf{B}=\mathbf{I}$, $\mathcal{U}^u$ represents interval uncertainty set.}

\subsection{Achievable Data Rate with Channel Gain Errors} We now derive the achievable data rate with channel errors. To this end,
let us rewrite the received signal in \eqref{rxedsig1} as in \eqref{eq:rxedsig_imperfectCSI},
shown at the top of the following page.
\begin{figure*}
\vspace{-7.049 mm}
{\color{black}
\begin{subequations}
\label{eq:rxedsig_imperfectCSI}
\begin{align}
y_{t} & =\sqrt{P_{t}}h_{t}x_{tt}+\sum_{j=t+1}^{u-1}h_{j}g_{jt}x_{tt}\sqrt{\beta_{jt}P_{t}}x_{jt}+h_{u}g_{ut}x_{tt}\sqrt{\beta_{ut}P_{t}}x_{ut}+\sum_{j=u+1}^{U}h_{j}g_{jt}x_{tt}\sqrt{\beta_{jt}P_{t}}x_{jt}+n_{t}\\
 & =\sqrt{P_{t}}(h_{t}+\Delta h_{t})x_{tt}+\sum_{j=t+1}^{u-1}(\bar{h}_{j}+\Delta h_{j})(\bar{g}_{jt}+\Delta{g}_{jt})x_{tt}\sqrt{\beta_{jt}P_{t}}x_{jt}+h_{u}g_{ut}x_{tt}\sqrt{\beta_{ut}P_{t}}x_{ut}\nonumber \\
& \quad+\sum_{j=u+1}^{U}h_{j}g_{jt}x_{tt}\sqrt{\beta_{jt}P_{t}}x_{jt}+n_{t}
 \\
 & =\sqrt{P_{t}}(h_{t}+\Delta h_{t})x_{tt}+\sum_{j=t+1}^{u-1}\bar{h}_{j}\bar{g}_{jt}x_{tt}\sqrt{\beta_{jt}P_{t}}x_{jt}+\sum_{j=t+1}^{u-1}(\bar{h}_{j}\Delta{g}_{jt}+\bar{g}_{jt}\Delta h_{j})x_{tt}\sqrt{\beta_{jt}P_{t}}x_{jt}\nonumber \\
 & \quad+\sum_{j=t+1}^{u-1}\Delta{g}_{jt}\Delta h_{j}x_{tt}\sqrt{\beta_{jt}P_{t}}x_{jt}+h_{u}g_{ut}x_{tt}\sqrt{\beta_{ut}P_{t}}x_{ut}+\sum_{j=u+1}^{U}h_{j}g_{jt}x_{tt}\sqrt{\beta_{jt}P_{t}}x_{jt}+n_{t}.
\end{align}
\end{subequations}
\hrule
\begin{equation}
R_{ut}=\begin{cases}
\log_{2}\left(1+\frac{|h_{u}|^{2}|P_{t}}{\sum_{j=u+1}^{U}|h_{j}|^{2}|g_{jt}|^{2}\beta_{jt}P_{t}+\sigma_{t}^{2}}\right) & u=t\\
\log_{2}\left(1+\frac{|h_{u}|^{2}|g_{ut}|^{2}\beta_{ut}P_{t}}{I_{SIC}+\sum_{j=u+1}^{U}|h_{j}|^{2}|g_{jt}|^{2}\beta_{jt}P_{t}+\sigma_{t}^{2}}\right) & u>t.
\end{cases}\label{eq:rate_imperfectCSI}
\end{equation}}
\end{figure*}

For users $u>t$, the SIC process can eliminate the term $\sum_{j=t+1}^{u-1}\bar{h}_{j}\bar{g}_{jt}x_{tt}\sqrt{\beta_{jt}P_{t}}x_{jt}$.
In practice, the channel errors are typically much smaller than the
nominal channel values. Thus, \textcolor{black}{in such circumstances,} it is reasonable to assume that the
term $\Delta{g}_{jt}\Delta h_{j}$ has a negligible impact on the achievable
data rate. \textcolor{black}{In general,} the achievable data rate for user $u$ in time slot $t$ \textcolor{black}{such that $u \geq t$}
takes the form as \eqref{eq:rate_imperfectCSI}, shown at the top
of the following page, where $I_{SIC}\triangleq\sum_{j=t}^{u-1}(|\bar{h}_{j}\Delta{g}_{jt}+\bar{g}_{jt}\Delta
h_{j}|^{2}+|\Delta{g}_{jt}\Delta h_{j}|^2)\beta_{jt}P_{t}$ represents the residual interference due to imperfect SIC, $\bar{g}_{tt}=1$ and $\Delta{g}_{tt}=0$.} 
\textcolor{black}{Note that the rate expression for $u=t$ follows from that of $u>t$ by observing that $I_{SIC}=0$ when $u=t$.}\par \textcolor{black}{It is pertinent to focus on $I_{SIC}$. Even in the presence of channel errors, the interference due to nominal channels known to the BS i.e., $\sum_{j=t+1}^{u-1}\bar{h}_{j}\bar{g}_{jt}x_{tt}\sqrt{\beta_{jt}P_{t}}x_{jt}$, is removed as a consequence of SIC. Hence, in $I_{SIC}$, only the terms that depend on CSI errors remain. Since CSI errors are generally quite small compared to the estimated values of the channels, in spirit of the work in \cite{JeffAnd_Imp_SIC}, $I_{SIC}$ can be approximated by the following simpler form. Specifically, if we assume that $\Pi_{jt}$ represents the error corresponding to the fraction of the uncancelled power of $j^{\text{th}}$ user in the $t^{\text{th}}$ time slot, $I_{SIC}$ can be represented as $I_{SIC}=\sum_{j=t}^{u-1} \Pi_{jt}\beta_{jt}P_{t}$. As noted in \cite{JeffAnd_Imp_SIC}, $\Pi_{jt}$ can be considered as the variance of the channel estimation error. In contrast to $I_{SIC}$, the interference from users $u+1$ to $U$ as shown as a sum of such terms in $R_{ut}$ is not processed in a manner similar to SIC. Therefore, we use the channel models defined in \eqref{actual_channels_1} and \eqref{actual_channels}, and proceed with the further developments in the sections to follow.}
\subsection{Problem Formulation}
H-NOMA is being proposed as an additional feature of existing TDMA based systems. 
An imperfect SIC causes additional interference, as reflected in the term $I_{SIC}$ in $R_{ut}$. %
In the presence of CSI errors, \textcolor{black}{it is not guaranteed that $R_{ut}$ meets the minimum data rate requirements for a user in a given time slot. It is, therefore, crucial that a threshold data rate is guaranteed in a time slot even in the presence of CSI errors. We adopt the worst-case robust optimization approach \cite{AharonNemirovskiGhaoui2009} to achieve this goal. The worst-case robust optimization, though fundamentally conservative in nature, ensures that even in the presence of significant channel estimation errors, the minimum threshold data rates can still be guaranteed for all users most of the time.}
We are interested in the following problem
\begin{subequations}\label{MProb}
\begin{align}
\mathop{{\rm minimize}}\limits _{P_u,\beta_{ut}} & \quad\sum_{u=1}^UP_u\\
{\rm subject~to} & \quad \sum_{t=1}^u R_{ut}\geq \mathcal{T},1\leq u\leq U,\forall\: \boldsymbol{\gamma}_{u}\in\mathcal{U}^{u},\boldsymbol{\gamma}_{ut}\in\mathcal{U}^{ut}\label{eq:rate}\\
 &\quad \beta_{ut}\leq 1,1\leq t\leq u-1,1\leq u \leq U
\end{align}
\end{subequations}
\textcolor{black}{where \eqref{eq:rate} ensures that a user $u$ completes its transmission by its allocated time slot and its total data rate across all time slots in which it transmits exceeds the threshold $\mathcal{T}$. Moreover, the worst-case robust design philosophy ensures that the threshold is met for all channel realizations in the given uncertainty sets.} Due to channel estimation errors, the problem in \eqref{MProb} is fundamentally different from that considered in \cite{DingHNoma}. The algorithmic approaches developed for the perfect CSI case in \cite{DingHNoma} are not applicable to \eqref{MProb} due to CSI errors.

\section{Proposed Robust Solution\label{PropSol}}
\subsection{Problem Reformulation}
The main problem considered in \eqref{MProb} is non-convex. The primary source of non-convexity in \eqref{MProb} stems from the data rate constraint in \eqref{eq:rate}, which remains non-convex even when there are no CSI errors. Moreover, it is observed from the additive channel estimation error model in \eqref{actual_channels_1} and \eqref{actual_channels}, that despite the finite number of optimization variables, due to the continuous nature of the uncertainty sets to which the perturbation vectors belong, see \eqref{uncertTyeI} and \eqref{uncertTyeII}, \eqref{eq:rate} makes the problem semi-infinite in nature. \textcolor{black}{In line with the arguments presented in \cite{DingHNoma}, in order to avoid the bilinear form of the decision variables, we make the substitutions $\beta_{ut}P_t=P_{ut}$ and $P_{uu}=P_u$}. %
Hence, first consider the equivalent formulation of \eqref{MProb} as
\begin{subequations}\label{MProb1}
\begin{align}
\mathop{{\rm minimize}}\limits _{P_{ut},t_{ut}} & \quad\sum_{u=1}^UP_u\\
{\rm subject~to} & \quad \sum_{t=1}^u t_{ut}\geq \mathcal{T},R_{ut}\geq t_{ut},1\leq u\leq U,\nonumber\\&\hspace{2.7cm}\forall\: \boldsymbol{\gamma}_{u}\in\mathcal{U}^{u},\boldsymbol{\gamma}_{ut}\in\mathcal{U}^{ut}\label{eq:rate1}\\
 &\quad P_{ut}\leq P_{tt},1\leq t\leq u-1,1\leq u \leq U\label{eq:powerConst}
\end{align}
\end{subequations}
where \textcolor{black}{the constraint in \eqref{eq:powerConst} is to ensure that $\beta_{ut}$ is no larger than unity, and the analysis variable $t_{ut}$ is implicitly non-negative for all $u$ and $t$. Further to this substitution, the residual interference due to imperfect SIC is taken as $I_{SIC}=\sum_{j=t+1}^{u-1} \Pi_{jt}P_{jt}$. To proceed ahead, we will take into account the more general expression of $R_{ut}$ in \eqref{eq:rate_imperfectCSI} which includes $I_{SIC}$.} Now consider the constraint involving $R_{ut}$ in \eqref{eq:rate1}, which can be rewritten as
\begin{align}
&\frac{|h_u|^2|g_{ut}|^2P_{ut}}{I_{SIC}+\sum_{j=u+1}^U|h_j|^2|g_{jt}|^2P_{jt}+\sigma_t^2}\geq \mathcal{\bar{T}},\nonumber\\&\hspace{1.1cm}1\leq t\leq u,1\leq u\leq U,\forall\: \boldsymbol{\gamma}_{u}\in\mathcal{U}^{u},\boldsymbol{\gamma}_{ut}\in\mathcal{U}^{ut}\label{sinrI} \\ &\Leftrightarrow |h_u|^2|g_{ut}|^2P_{ut}\geq \nonumber\\&\mathcal{\bar{T}}(I_{SIC}+\sum_{j=u+1}^U|h_j|^2|g_{jt}|^2P_{jt}+\sigma_t^2),1\leq t\leq u,\nonumber\\&1\leq u\leq U,\forall\: \boldsymbol{\gamma}_{u}\in\mathcal{U}^{u},\boldsymbol{\gamma}_{ut}\in\mathcal{U}^{ut}\label{sinrIa}\\&\Leftrightarrow\begin{cases} |h_u|^2|g_{ut}|^2\geq \frac{\epsilon_{ut}^2}{P_{ut}},1\leq t\leq u,1\leq u\leq U,\\\forall\: \boldsymbol{\gamma}_{u}\in\mathcal{U}^{u},\boldsymbol{\gamma}_{ut}\in\mathcal{U}^{ut}\\\epsilon_{ut}^2\geq\mathcal{\bar{T}}(\sum_{j=t+1}^{u-1} \Pi_{jt}P_{jt}+\\\sum_{j=u+1}^U|h_j|^2|g_{jt}|^2P_{jt}+\sigma_t^2),1\leq t\leq u,\\1\leq u\leq U,\forall\: \boldsymbol{\gamma}_{u}\in\mathcal{U}^{u},\boldsymbol{\gamma}_{ut}\in\mathcal{U}^{ut}\label{sinrII}\end{cases}
\end{align}
where $\mathcal{\bar{T}}=2^{t_{ut}}-1$, and the auxiliary variables $\epsilon_{ut}^2$ are introduced to obtain the equivalent transformation in \eqref{sinrII}. 
\begin{remark}\label{Rem_bilinear}
In its present form, the equivalent formulation in \eqref{sinrII} is not yet computationally tractable. Among other issues, a major source of concern is the product of erroneous channel gains of the form $|h_u|^2|g_{ut}|^2$. Traditionally, channel estimation errors have not been addressed for such product forms, for instance, \cite{HanifTran_Robust} and the references therein. Initially, we obtain a tractable version of the first constraint and later the second constraint in \eqref{sinrII}. To the best of authors' knowledge, the adopted developments are not known in the context of NOMA systems. 
\end{remark}
\begin{thm}
\label{thm1:1stApp} For the uncertainty sets defined in \eqref{uncertTyeI} and \eqref{uncertTyeII}, consider the constraint given below 
\begin{align}
\min_{\boldsymbol{\gamma}_{u}\in\mathcal{U}^{u},\boldsymbol{\gamma}_{ut}\in\mathcal{U}^{ut}}|h_u|^2|g_{ut}|^2\geq \frac{\epsilon_{ut}^2}{P_{ut}},1\leq t\leq u,1\leq u\leq U.\label{sinrIIeqathm0}
\end{align}
It can be shown that the following set of constraints is equivalent to \eqref{sinrIIeqathm0}
\begin{align}
\exists \boldsymbol{\lambda}_u:|\bar{h}_u|^2|\bar{g}_{ut}|^2+\begin{cases}-\rho\|\mathbf{b}_{ut}\|-\boldsymbol{\lambda}_u^T\mathbf{l}\geq\frac{\epsilon_{ut}^2}{P_{ut}},1\leq t\leq u,\\\hspace{3.5cm}1\leq u\leq U\\-a_u^w-\rho\|\mathbf{q}_{ut}^w\|_2+\boldsymbol{\lambda}_u^T\mathbf{a}^w\geq 0,\\1\leq w\leq L,\boldsymbol{\lambda}_u\geq 0,1\leq t\leq u,\\\hspace{3.5cm}1\leq u\leq U.\end{cases}\label{sinrIIeqathm1}
\end{align}
\end{thm}
\begin{IEEEproof}
The proof of the theorem is based on the concepts of the worst-case robust optimization and Lagrange duality theory. The details of the proof are presented in Appendix~\ref{Proof_thm1}. 
\end{IEEEproof}
Let $f(\epsilon_{ut},P_{ut})\triangleq\frac{\epsilon_{ut}^2}{P_{ut}}$. The convexity of $f(\epsilon_{ut},P_{ut})$ follows from the arguments given in \cite[Sec. 3.2.6]{BoydCVX2004}. %
Consider next the second inequality in \eqref{sinrII} which can be rewritten as the system given below
\begin{align}
&\epsilon_{ut}^2\geq f^\prime(t_{ut},P_{jt},\zeta_{jt}),1\leq t\leq u,1\leq u\leq U\label{secI_sinrII}\\&|h_j|^2|g_{jt}|^2\leq \frac{\zeta_{jt}^2}{P_{jt}},1\leq t\leq u,u+1\leq j\leq U,\nonumber\\&\hspace{4.0cm}\forall\: \boldsymbol{\gamma}_{u}\in\mathcal{U}^{u},\boldsymbol{\gamma}_{ut}\in\mathcal{U}^{ut}\label{secIII_sinrII}%
\end{align}
where $f^\prime(t_{ut},P_{jt},\zeta_{jt})\triangleq \mathcal{\bar{T}}(\sum_{j=t}^{u-1} \Pi_{jt}P_{jt}+\sum_{j=u+1}^U\zeta_{jt}^2+\sigma_t^2)$. The constraints \eqref{secI_sinrII} %
and \eqref{secIII_sinrII} are all non-convex and intractable. In light of the observation made above, the right-hand side of the inequality in \eqref{secIII_sinrII} represents the convex quadratic over linear function $f(\zeta_{jt},P_{jt})=\frac{\zeta_{ij}^2}{P_{jt}}$. %
Now we use techniques from the majorization-minimization framework \cite{PalomarMM} to bound the non-convex functions in \eqref{secI_sinrII} and \eqref{secIII_sinrII} using convex surrogates.
\begin{lema}\label{cvx_tngnt_prop}
The functions $\epsilon_{ut}^2, f(\vartheta,\varphi)$
and $f^\prime(t_{ut},P_{jt},\zeta_{jt})$, where $f(\vartheta,\varphi)=\vartheta^2/\varphi$ is a convex quadratic over linear function can be bounded as follows
\begin{enumerate}
\item $\epsilon_{ut}^2\geq \epsilon_{ut}^n(2\epsilon_{ut}-\epsilon_{ut}^n)$\label{lb_1}
\item $f(\vartheta,\varphi)\geq g(\vartheta,\varphi;\vartheta^{n},\varphi^{n})$\label{lb_2n}\item $f^\prime(t_{ut},P_{jt},\zeta_{jt})\leq \frac{\lambda_{ut}}{2}\mathcal{\bar{T}}^2+\frac{1}{2\lambda_{ut}}\bar{g}^2(P_{jt},\zeta_{jt},\sigma_t)$\label{ub}
\end{enumerate}
where $g(\vartheta,\varphi;\vartheta^{n},\varphi^{n})\triangleq \frac{\vartheta^2}{\varphi}\big|_{n}+\frac{2\vartheta}{\varphi}\big|_{n}(\vartheta-\vartheta^{n})-\frac{\vartheta^2}{\varphi^2}\big|_{n}(\varphi-\varphi^{n}), \bar{g}(P_{jt},\zeta_{jt},\sigma_t)\triangleq (\sum_{j=t}^{u-1} \Pi_{jt}P_{jt}+\sum_{j=u+1}^U\zeta_{jt}^2+\sigma_t^2),\lambda_{ut}=\frac{\bar{g}(P_{jt},\zeta_{jt},\sigma_t)}{\mathcal{\bar{T}}}|_n$, the notation $|_{n}$ denotes the value of the preceding function at $n$,  and when $n$ is used as a superscript on a variable, it represents its value at $n$. It is noted that all the above bounds are held with equality when the variables in each function are replaced with their values at $n$. The bound on $f(\zeta_{jt},P_{jt})=\frac{\zeta_{jt}^2}{P_{jt}}$ simply follow from the bound on $f(\vartheta,\varphi)$ in 2).
\end{lema}
\begin{IEEEproof}
For completeness, we briefly outline the proof of these bounds. The lower bounds in \ref{lb_1}) and \ref{lb_2n})  presented above follow from the observation that the convex functions $\epsilon_{ut}^2$ and $f(\vartheta,\varphi)$ are always above their first-order Taylor series approximation \cite{BoydCVX2004}. The case of equality in which the variables are replaced with their values at $n$ is straightforward to observe. The bound in \ref{ub}) follows from the fact that for positive $a$ and $b$, using the arithmetic-geometric mean inequality, $ab\leq \frac{\lambda}{2}a^2+\frac{1}{2\lambda}b^2$, where $\lambda=b/a$ leads to equality.
\end{IEEEproof}
Hence, now using the bounds given in Lemma \ref{cvx_tngnt_prop}, the tractable version of \eqref{secI_sinrII} becomes 
\begin{align}
&\epsilon_{ut}^n(2\epsilon_{ut}-\epsilon_{ut}^n)\geq \frac{\lambda_{ut}}{2}\mathcal{\bar{T}}^2+\frac{1}{2\lambda_{ut}}\bar{g}^2(P_{jt},\zeta_{jt},\sigma_t),\nonumber\\&\hspace{3.5 cm}1\leq t\leq u,1\leq u\leq U.\label{sinr0eqa}
\end{align}
In a similar way, the robust and tractable version of \eqref{secIII_sinrII} follows from Theorem \ref{thm1:1stApp} 
\begin{align}
&\max_{\boldsymbol{\gamma}_{u}\in\mathcal{U}^{u},\boldsymbol{\gamma}_{ut}\in\mathcal{U}^{ut}}|h_j|^2|g_{jt}|^2\leq g(\zeta_{jt},P_{jt};\zeta_{jt}^{n},P_{jt}^{n}),1\leq t\leq u,\nonumber\\&\hspace{5.1 cm}u+1\leq j\leq U\nonumber\\&\hspace{-.1 cm}\Leftrightarrow
\exists \boldsymbol{\omega}_j:|\bar{h}_j|^2|\bar{g}_{jt}|^2+
\begin{cases}\rho\|\mathbf{b}_{jt}\|+\boldsymbol{\omega}_j^T\mathbf{u}\leq g(\zeta_{jt},P_{jt};\zeta_{jt}^{n},P_{jt}^{n}),\\1\leq t\leq u,
u+1\leq j\leq U\\-a_{j}^w-\rho\|\mathbf{q}_{jt}^w\|_2+\boldsymbol{\omega}_j^T\mathbf{b}^w\geq 0,\\\boldsymbol{\omega}_j\geq 0,1\leq w\leq L, 1\leq t\leq u,\\ \hspace{2cm}
u+1 \leq j \leq U.
\end{cases}\label{sinrIIeqa}
\end{align}
Note that unlike dualizing a minimization problem, \eqref{sinrIIeqa} is obtained by dualizing a maximization problem on similar grounds as described in Appendix~\ref{Proof_thm1}. \par Each non-convex constraint in the original problem formulation has now been written in a tractable and convex form. Therefore, the convex approximation of the optimization problem in \eqref{MProb} is as shown in \eqref{MProb_Approx} at the top of the next page,
\begin{figure*}
\begin{subequations}\label{MProb_Approx}
\begin{align}
\mathop{{\rm minimize}}\limits _{P_{ut},t_{ut},\boldsymbol{\lambda}_u,\boldsymbol{\zeta},\epsilon_{ut},\boldsymbol{\omega}_j}&\quad\sum_{u=1}^UP_{uu}\label{sinrIIeqaFinPro_obj}\\ 
{\rm subject~to} &\sum_{t=1}^u t_{ut}\geq \mathcal{T}, 1\leq u \leq U \label{sinrIIeqaFinPro0} \\ & \hspace{-0.1cm} |\bar{h}_u|^2|\bar{g}_{ut}|^2-\rho\|\mathbf{b}_{ut}\|-\boldsymbol{\lambda}_u^T\mathbf{l} \geq\frac{\epsilon_{ut}^2}{P_{ut}},1\leq t\leq u,1\leq u\leq U\label{sinrIIeqaFinPro1}\\& -a_u^w-\rho\|\mathbf{q}_{ut}^w\|_2+\boldsymbol{\lambda}_u^T\mathbf{a}^w\geq 0,1\leq w\leq L,\boldsymbol{\lambda}_u\geq 0,1\leq t\leq u,1\leq u\leq U\label{sinrIIeqaFinPro2}\\
 & \hspace{0 mm}\epsilon_{ut}^n(2\epsilon_{ut}-\epsilon_{ut}^n)\geq \frac{\lambda_{ut}}{2}\mathcal{\bar{T}}^2+\frac{1}{2\lambda_{ut}}\bar{g}^2(P_{jt},\zeta_{jt},\sigma_t),1\leq t\leq u,1\leq u\leq U\label{secI_sinrIIFinPro3}\\&\hspace{0 cm}|\bar{h}_j|^2|\bar{g}_{jt}|^2+\rho\|\mathbf{b}_{jt}\|+\boldsymbol{\omega}_j^T\mathbf{u}\leq g(\zeta_{jt},P_{jt};\zeta_{jt}^{n},P_{jt}^{n}),1\leq t\leq u,
u+1\leq j\leq U,\forall u\in\mathcal{U}\label{sinrIeqaFinPro6}\\&-a_{j}^w-\rho\|\mathbf{q}_{jt}^w\|_2+\boldsymbol{\omega}_j^T\mathbf{b}^w\geq 0,\boldsymbol{\omega}_j\geq 0,1\leq w\leq L,1\leq t\leq u,
u+1\leq j\leq U,\forall u\in\mathcal{U}\label{sinrIIeqaFinPro7}
\\&\hspace{0 cm}P_{ut}\leq P_{tt},1\leq t\leq u-1,1\leq u\leq U. \label{sinrIIeqaFinPro8}
\end{align}
\end{subequations}
\hrule
\end{figure*}
where the dual variables $\boldsymbol{\lambda}_u$ and $\boldsymbol{\omega}_j$ in \eqref{MProb_Approx} are all non-negative. 
\begin{lema}\label{subset_feasibility}
The feasibility set of the formulation in \eqref{MProb_Approx} is a subset of the original problem in \eqref{MProb}.
\end{lema}
\begin{IEEEproof}
The main ingredient of the proof is the set of inequalities in Lemma \ref{cvx_tngnt_prop}. To see this, we consider the constraint set in \eqref{secI_sinrIIFinPro3}. For all $1\leq t\leq u,1\leq u \leq U$, consider a set of variables $\epsilon_{ut},t_{ut}$ and $\zeta_{jt}$ that satisfy the inequality in \eqref{secI_sinrIIFinPro3}. Now the same set of variables will also satisfy the inequality in \eqref{secI_sinrII} due to the relation presented in 1) and 2) of Lemma \ref{cvx_tngnt_prop}. The same argument can also be extended to the variables that satisfy the inequality in \eqref{sinrIeqaFinPro6}. Therefore, the set of variables that satisfies \eqref{MProb_Approx} is also feasible in the original formulation without any approximating functions. To conclude, the feasible set of \eqref{MProb_Approx} is a subset of the feasibility set of \eqref{MProb1}, and hence that of \eqref{MProb}.
\end{IEEEproof}
The above lemma shows that \eqref{MProb_Approx} is a conservative approximation of the original problem in \eqref{MProb}. This can lead to difficulties in obtaining an initial feasible point. In our numerical investigations, we particularly note that the constraint in \eqref{sinrIeqaFinPro6} contributes the most to the conservatism of \eqref{MProb_Approx}. Therefore, we add slack variables to this constraint and obtain a penalized form (see \cite{Inosha_Topology}) of problem \eqref{MProb_Approx}. Although this approach is similar to the one considered in \cite{BoydDiffCvx}, it fundamentally differs from the existing one in terms of usage. More precisely, \cite{BoydDiffCvx} develops the penalty convex concave procedure specifically for the difference of convex (DC) programming problems. In our case, meanwhile, we obtain a penalized formulation for a general convex optimization problem. Consequently, the problem in \eqref{MProb_Approx}, with penalized objective, and the updated constraints in \eqref{sinrIeqaFinPro6}, takes the following form
\begin{subequations}\label{MProb_Approx_sl}
\begin{align}
\mathop{{\rm minimize}}\limits _{\substack{P_{ut},t_{ut},\boldsymbol{\lambda}_u,\boldsymbol{\zeta},\\\epsilon_{ut},\boldsymbol{\omega}_j,s_{jt}}}  &\quad\sum_{u=1}^UP_{uu}+\tau_n\sum_{j,t}s_{jt}\label{sinrIIeqaFinPro_obj_pen}\\
{\rm subject~to}&\hspace{6 mm} \eqref{sinrIIeqaFinPro0}, \eqref{sinrIIeqaFinPro1},\eqref{sinrIIeqaFinPro2},\eqref{secI_sinrIIFinPro3},\eqref{sinrIIeqaFinPro7},\eqref{sinrIIeqaFinPro8}\nonumber\\&\hspace{-2cm}|\bar{h}_j|^2|\bar{g}_{jt}|^2+\rho\|\mathbf{b}_{jt}\|+\boldsymbol{\omega}_j^T\mathbf{u}\leq g(\zeta_{jt},P_{jt};\zeta_{jt}^{n},P_{jt}^{n})+s_{jt},\nonumber\\&1\leq t\leq u, 
u+1\leq j\leq U,\forall u \in\mathcal{U}\label{sinrIeqaFinPro6_sl}
\end{align}
\end{subequations}
where $s_{jt}$ are the non-negative slack variables, and $\tau_n$ is constant. The constant $\tau_n$ is varied so that, at the cost of violating the constraints, a feasible objective value is returned. The sum of violations is penalized in the objective, so that the violated constraints are as small as possible.
\subsection{Algorithmic Solution}
We develop an iterative procedure to solve the problem in \eqref{MProb_Approx_sl}. We take advantage of the minorization-majorization (MM) framework in which the objective is progressively improved after each iteration of the algorithm \cite{HunterMM}. It is important to note that the MM approach is a general method to develop optimization algorithms that can be applied to a wide variety of problems \cite{PalomarMM}. 
\begin{algorithm}
\caption{Robust H-NOMA power minimization algorithm}
\label{alg:RHNOMA}
\KwIn{\text{Nominal channels, parameters of the system model}\newline \text{and the uncertainty sets}, $P_{ut}^{0}, \epsilon_{ut}^{0}, \zeta_{jt}^{0},\lambda_{ut}^0,\forall u,t,j, \tau_0>0,\xi< 1$}
\KwOut{ $P_{ut}^{\star},t_{ut}^\star,\boldsymbol{\lambda}_u^{\star},\boldsymbol{\zeta}^{\star},\epsilon_{ut}^{\star},\boldsymbol{\omega}_j^{\star},s_{jt}^\star$}
$n\leftarrow0$\;
\Repeat{convergence }{
Solve~\eqref{MProb_Approx_sl} and let $P_{ut}^{\star}$, $\epsilon_{ut}^{\star}$, $\zeta_{jt}^{\star}$ \ denote the optimal decision variables\;
Update: $P_{ut}^{n+1}\leftarrow P_{ut}^{\star},\epsilon_{ut}^{n+1}\leftarrow \epsilon_{ut}^{\star},\zeta_{jt}^{n+1}\leftarrow\zeta_{jt}^{\star},\lambda_{ut}^{n+1}\leftarrow\frac{\bar{g}^2(P_{jt},\zeta_{jt},\sigma_t)}{\mathcal{\bar{T}}}|_n,\tau^{n+1}\leftarrow \min\{\frac{\tau^n}{\xi},\tau_{\max}\}$\;
$n\leftarrow n+1$}
\end{algorithm}
An algorithm for robust power minimization in H-NOMA uplink with users equipped with BackCom circuits is outlined in \textbf{Algorithm \ref{alg:RHNOMA}}. In our algorithm, the index $n$ is used to represent the iteration of the algorithm, and it corresponds to the value of the functions or variables in iteration $n$ of the procedure. The weight of the penalty term $\tau^n$ is updated in each iteration using a suitable value of $\xi$. It is emphasized that the weights of the penalty term $\tau_n$, in \textbf{Algorithm \ref{alg:RHNOMA}}, are updated in a manner similar to that used for DC programs in \cite{BoydDiffCvx}. In the proposed algorithm, it is seen that in the approximation of $f(\zeta_{jt},P_{jt})$, a divide-by-zero error may occur if $P_{ut}^n=0$ for some $u,t$ and $n$. To circumvent this situation and make the algorithm more robust, either a small positive constant can be added to the denominator of the update or additional constraints, which ensure that the power variables are greater than or equal to a suitably chosen positive constant can be incorporated into the main problem. The reflection coefficients $\beta_{ut}$ follow from $P_{ut}^{\star}=\beta_{ut}P_t^{\star}$.

\begin{lema}\label{lems:convergence}
The sequence of objective values returned by \textbf{Algorithm \ref{alg:RHNOMA}} is guaranteed to converge to the Karush-Kuhn-Tucker (KKT) point of \eqref{MProb}.
\end{lema}
\begin{IEEEproof}
The proof of this fact follows similar arguments as given in \cite{BoydDiffCvx,HunterMM,PalomarMM,MarksInnerApprox}, and is briefly presented in the following to make the article self-contained. Let $\mathcal{O}_n,\mathcal{V}_n$ and $\mathcal{F}_n$ be the objective, the set of decision variables and the feasible set in the $n^{\text{th}}$ iteration of \textbf{Algorithm \ref{alg:RHNOMA}}. As a consequence of Lemma \ref{subset_feasibility}, it is easy to observe that $\mathcal{V}_n$ belongs to both $\mathcal{F}_n$ and $\mathcal{F}_{n+1}$. Then, the optimality of $\mathcal{V}_{n+1}$ in the $n+1^{\text{st}}$ iteration of the \textbf{Algorithm \ref{alg:RHNOMA}} implies $\mathcal{O}_{n+1}\leq \mathcal{O}_{n}$. Guaranteed convergence follows from the compactness of the feasible region of \eqref{MProb}. It is important to note that if the slack variables have not diminished to zero, it is not necessary that the convergence point is also feasible to the original problem \cite{BoydDiffCvx}. When the algorithm takes finite runs to converge and slack variables tend, the convergence to the KKT point follows similar arguments as given \cite{Thi_1,MarksInnerApprox,KumarNOMA}.
\end{IEEEproof}
\begin{remark}\label{KKT_global}
Note that the convergence of \textbf{Algorithm \ref{alg:RHNOMA}} to the KKT point does not necessarily imply its global optimality. The globally optimal solution can be obtained with certainty by using purely non-convex programming techniques like branch and bound / branch and cut methods or their appropriate combination with convex approximation methods such as DC programming \cite{Thi_2}, which is beyond the scope of this work. Nevertheless, in Sec. \ref{Res} we gauge the performance of our \textbf{Algorithm \ref{alg:RHNOMA}} and ascertain its superiority by benchmarking it against solutions of well known existing scenarios.
\end{remark}
\subsection{Complexity of \textbf{Algorithm \ref{alg:RHNOMA}}}
\textcolor{black}{The worst-case complexity of \textbf{Algorithm \ref{alg:RHNOMA}} is predominantly affected by its per iteration complexity. More specifically, the complexity of solving the convex problem \eqref{MProb_Approx} is required to characterize the overall complexity of our algorithm. The convex formulation in \eqref{MProb_Approx} consists of linear and conic quadratic constraints. It is well known that although the complexity of conic quadratic constraints is slightly higher than that of linear constraints, both can be solved with polynomial complexity using interior point methods \cite[Ch. 6]{AharonNemirovski2001}. Now using the results in \cite[Ch. 6]{AharonNemirovski2001}, the worst-case complexity estimate of \eqref{MProb_Approx} can be obtained as $O(\max(U^{3.5},UL(L^2+U^2),U^3L^{3.5},U^4,(U^2+L+2)^{1.5}L^2,(LU^2+L)^{1.5}L^2,U^3))$. To elaborate on this expression, assume that $U=L$, so that the complexity expression simplifies to $O(U^{6.5})$. Despite being polynomial, this theoretical complexity is rather high. However, the off-the-shelf implementations \cite{mosek1} of interior point methods exploit factors like sparsity resulting in a much lower real time worst-case complexity. For a given initialization, the overall complexity of \textbf{Algorithm \ref{alg:RHNOMA}} is dependent on the number of iterations it takes to converge. The total number of iterations is either set as a fixed number or for all practical purposes this number is a moderate constant for widely used convergence criteria.}
\section{Numerical Results}
\label{Res}
\begin{figure}[tb]
		\centering
        \begin{subfigure}{0.5\textwidth}\centering
\includegraphics[width=1\columnwidth]{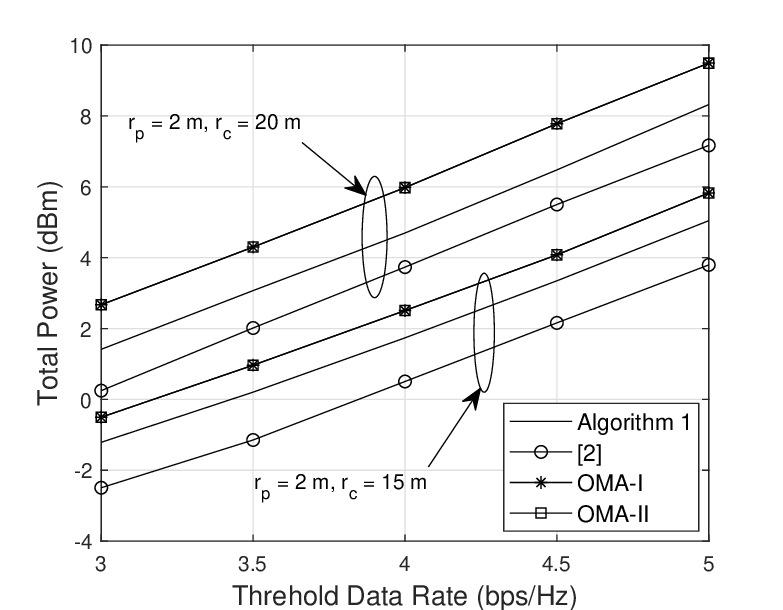}
		\caption{Total power as a function of threshold data rate. The uncertainty parameter $\rho$ is taken as 0.025.}
		\label{fig:tauvsP_rho_0.05}
        \end{subfigure}
        \begin{subfigure}{0.5\textwidth}\centering
        \includegraphics[width=1\columnwidth]{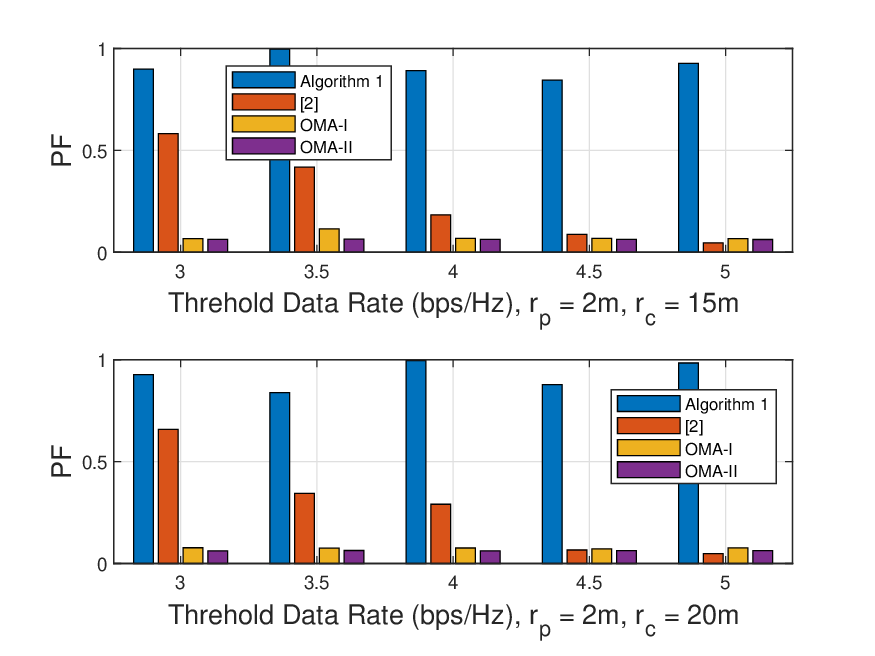}
		\caption{PF versus threshold data rate with uncertainty parameter $\rho$ taken as 0.025.}
		\label{fig:tauvsPF_rho_0.05}
        \end{subfigure}
        \caption{Variation of total consumed power and PF with threshold data rate. The uncertainty parameter $\rho$ is set as 0.025. }\label{fig:PowPFvstau_rho_0.05}
	\end{figure}
    \begin{figure}[tb]
		\centering
\includegraphics[width=1\columnwidth]{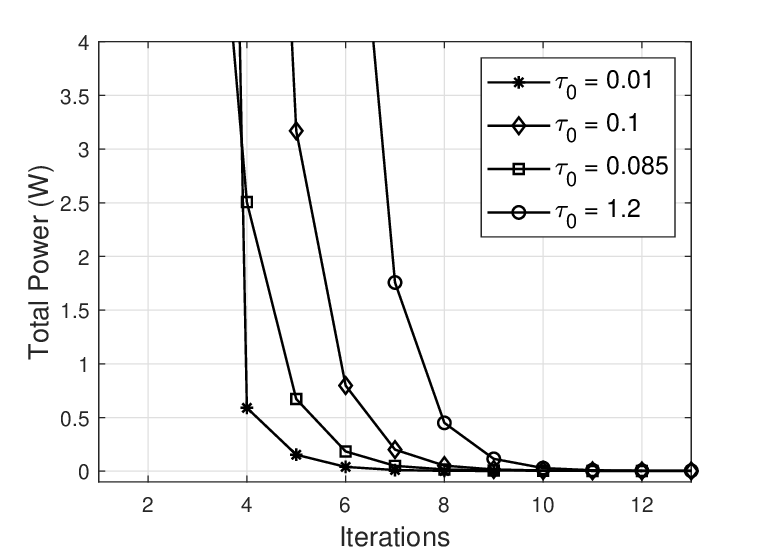}
		\caption{Convergence behaviour of \textbf{Algorithm \ref{alg:RHNOMA}} for different $\tau_0$. The parameter $\rho$ is fixed as $0.025$, $r_p=2$ m and $r_c=15$ m.}
		\label{fig: PowVsIter}
	\end{figure}
     \begin{figure}[tb]
		\centering
\includegraphics[width=1\columnwidth]{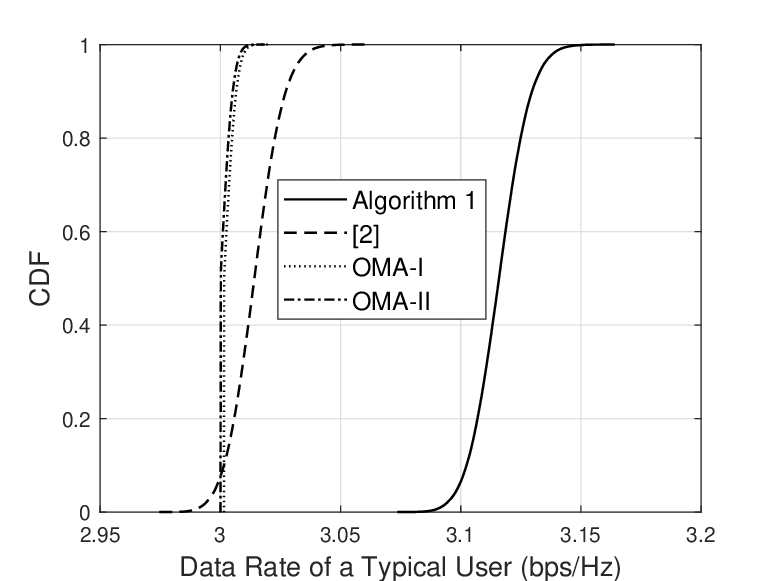}
		\caption{CDF plot of data rate (bps/Hz) of a typical user. We take $\rho=0.025$, $\mathcal{T}=3$ bps/Hz, $r_p=2$ m and $r_c=15$ m.}
		\label{fig: CDF_DataRate_User}
	\end{figure}
    \begin{figure}[tb]
		\centering
\includegraphics[width=1\columnwidth]{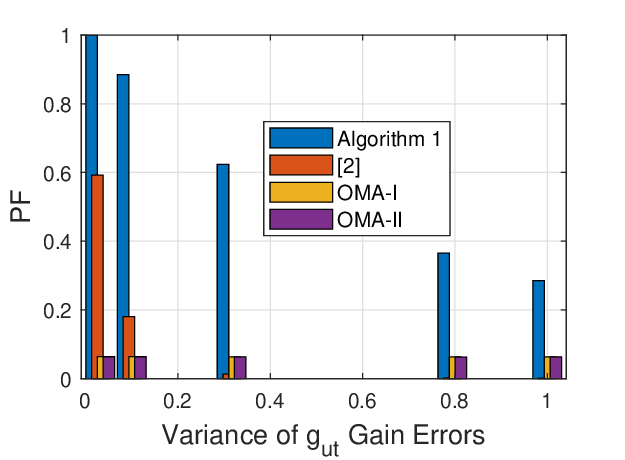}
		\caption{PF vs variance of channel gain errors of $g_{ut}$. The magnitude of channel estimation error of $h_{u}$ is kept fixed at $10^{-2.5}$. The threshold data rate $\mathcal{T}=4$ bps/Hz, $r_p=2$ m and $r_c=15$ m.}
		\label{fig: PFvsg_ut_errs}
	\end{figure}
\begin{figure}[tb]
		\centering
\includegraphics[width=1\columnwidth]{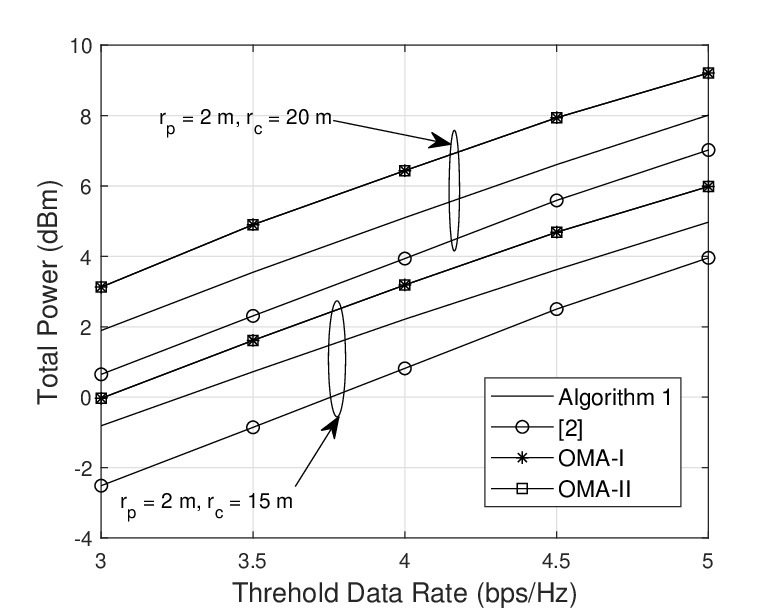}
		\caption{Total power as a function of threshold data rate, $\mathcal{T}$. The uncertainty parameter $\rho$ is set as 0.02 with dense $\mathbf{A}$ and $\mathbf{B}$.}
		\label{fig: tau_vs_TPow_Full_AB}
	\end{figure}
To evaluate the performance of the proposed solution based on \textbf{Algorithm \ref{alg:RHNOMA}} to the main problem in \eqref{alg:RHNOMA}, we consider several particular cases. In a similar spirit to \cite{DingHNoma}, we assume that $U$ users are uniformly distributed in a square region with side length being represented as $r_p$ m. The BS station is assumed to be located at the origin and the centre of the square region is positioned at coordinates $(r_c,r_c)$ m. The channels between the users and the BS are assumed to be Rayleigh faded and those between the users are assumed to be Rician faded with Rician factor $K$ dB. Both categories of channels are assumed to encounter distance dependent path loss. Rician factor and path loss exponents of Rayliegh and Rician channels are taken as 10 dB, 3 and 2, respectively. Regarding the modeling of CSI errors, without loss of generality, we generate matrices $\mathbf{A}=\mathbf{\bar{A}}^T\mathbf{\bar{A}},\mathbf{B}=\mathbf{\bar{B}}^T\mathbf{\bar{B}}$, where the entries of $\mathbf{\bar{A}}$ and $\mathbf{\bar{B}}$ are uniformly distributed in $(0.5,20)$. For the positive definite matrices $\mathbf{A}$ and $\mathbf{B}$, in the simulations, we consider the corresponding diagonal matrices with eigenvalues of $\mathbf{A}$ and $\mathbf{B}$ on their leading diagonals. Unless otherwise stated, the vector $\mathbf{l}$ is uniformly distributed in the interval $(0.5,10)$, $\mathbf{u}$ is at least five times $\mathbf{l}$ and $\Pi_{jt}$ are all uniformly distributed in $(0,0.1)$. It is assumed that $L=3, U=4$ and $\tau_{\max}=150$. The perturbation vectors $\boldsymbol{\alpha}_u$ and $\boldsymbol{\kappa}_{ut}$ are also uniformly distributed in $(0.05,2)$. Unless otherwise stated, the noise variance is taken as $\sigma_t^2=10^{-9}$. In \textbf{Algorithm \ref{alg:RHNOMA}}, convergence is achieved when the difference between two successive objective values is less than or equal to $10^{-3}$. We represent the results of the proposed approach to minimize power with CSI errors in BackCom H-NOMA as ``Algorithm 1''. For comparison, we also include results for BackCom H-NOMA with nominal channels only similar to the settings of \cite{DingHNoma}. The results of OMA with uncertainty sets \eqref{uncertTyeI} and \eqref{uncertTyeII} is shown as ``OMA-I'' and ``OMA-II'', respectively. The presented simulation results are obtained via CVX \cite{cvx} using MOSEK \cite{mosek1} as the internal solver. \par The measure of ``probability of feasibility (PF)'' of \eqref{MProb} in the presence of channel gain errors is considered. More precisely, for a given set of nominal channels, we solve \eqref{MProb_Approx_sl} using \textbf{Algorithm \ref{alg:RHNOMA}}. Then for several thousand realizations of normally distributed errors in channel gains, the fraction of channel instances is determined for which the minimum data rate based quality of service (QoS) constraints of all $U$ users are satisfied. Thus, the degree of robustness of a particular scheme against channel gain errors is gauged. One can observe that PF is just a particular value of the probability mass function (pmf) of a discrete random variable that represents the number of users that meet the QoS criterion. Clearly, a lower power consumption, together with a higher PF, represents the ideal scenario. A multiple access technique with significantly low PF is useless no matter how small the amount of power it consumes. \par
In Fig. \ref{fig:PowPFvstau_rho_0.05}(\subref{fig:tauvsP_rho_0.05}), the variation of the total power with the threshold data rate, $\mathcal{T}$, is studied. The robust BackCom H-NOMA scheme based on \textbf{Algorithm \ref{alg:RHNOMA}} consumes more power than the one based on nominal channels \cite{DingHNoma}, but is always more power efficient than the OMA scenarios. Moreover, the power used by both OMA-I and OMA-II is similar. The power consumption of the robust scheme is higher than that for the nominal channels based transmission \cite{DingHNoma}, since we use the worst-case robust optimization technique which renders its feasible set to be a subset of the original problem, as deducible from Lemma \ref{subset_feasibility}. It is pertinent to note that with increase in $r_c$, the total power requirement of all schemes is enhanced. This observation follows from the fact that increasing $r_c$ results in greater path loss and hence more power is needed to meet the minimum data rate threshold of each user.\par Fig. \ref{fig:PowPFvstau_rho_0.05}(\subref{fig:tauvsPF_rho_0.05}) complements Fig. \ref{fig:PowPFvstau_rho_0.05}(\subref{fig:tauvsP_rho_0.05}). PF matters the most when a given design is tested against channel gain errors. For each multiple access technique and a typical set of nominal channels, we generate several thousand zero mean Gaussain distributed channel gain errors with certain variance, and determine the probability that all $U$ users meet the QoS constraints. It is seen in  Fig. \ref{fig:PowPFvstau_rho_0.05}(\subref{fig:tauvsPF_rho_0.05}) that the proposed robust design outperforms all other multiple access techniques by a considerable  margin. Specifically, when $r_c=20$ m, at a data rate threshold value of 5 bps/Hz, \textbf{Algorithm \ref{alg:RHNOMA}} ensures the QoS constraints of all users are satisfied nearly with probability 1. In contrast, PF is below 10\% for all other schemes.\par Fig. \ref{fig: PowVsIter} shows the convergence behaviour of the iterative \textbf{Algorithm \ref{alg:RHNOMA}} with different initial weights of the penalty term. It is seen in the figure that irrespective of starting weights, the algorithm converges to the same point. However, the convergence is quickest for the smallest $\tau_0$. By initially using more weight on the penalty term, the iterative procedure needs more runs to minimize the penalty term in the objective. Although not observed in our settings, it is quite possible that a large initial point value $\tau_0$ can quickly make $\tau^n=\tau_{\max}$ for some $n$. Consequently, the objective value may converge before sufficiently minimizing the violations in constraints involving slack variables. \par Fig. \ref{fig: CDF_DataRate_User} shows the cumulative distribution functions (CDFs) of the achievable data rate of one of the $U$ users with normally distributed channel estimation errors for different multiple access strategies. The CDFs show that for the given threshold data rate of $3$ bps/Hz, except for our scheme based on \textbf{Algorithm \ref{alg:RHNOMA}}, all other multiple access methods remain below this threshold rate for a substantial amount of time. Even if for some user, if some data rate threshold values are exceeded, the overall problem remains infeasible since for that to happen all $U$ users should meet the threshold constraints simultaneously. Remarkably, CDFs of both types of OMA schemes show that the achievable rates of the OMA user nearly always remains below the threshold of 3 bps/Hz.\par
In Fig. \ref{fig: PFvsg_ut_errs}, we study the impact of CSI errors for different multiple access techniques. We design each system for a threshold data rate of $4$ bps/Hz using the standard parameters considered in this section. Keeping the variance of channel gain errors fixed for all $h_{u}$, the magnitude of channel gain errors of $g_{ut}$ is varied, and PF is calculated. It is seen that as the strength of channel gain errors increases, the PF decreases in all schemes. However, despite the worst channel errors, the PF of the proposed scheme remains high compared to all other multiple access techniques. It shows that, while retaining its power efficiency, the robust H-NOMA based on \textbf{Algorithm \ref{alg:RHNOMA}} ensures resilience against channel estimation errors. \par
Fig. \ref{fig: tau_vs_TPow_Full_AB} compares all multiple access schemes when the uncertainty matrices $\mathbf{A}$ and $\mathbf{B}$ in \eqref{uncertTyeI} are dense and fully used, i.e, diagonal matrices with eigenvalues of $\mathbf{A}$ and $\mathbf{B}$ are not considered. Remarkably, it is seen that \textbf{Algorithm \ref{alg:RHNOMA}} outperforms other approaches for both cases $r_c=15$ m and $r_c=20$ m. Therefore, at least for the parameters considered in this section, the proposed method is not only power efficient but also promises a higher value of PF even if we do not restrict $\mathbf{A}$ and $\mathbf{B}$ to be diagonal.
\section{Conclusion}
\label{Conc} 
In this paper, we have studied power minimization in the uplink of a BackCom H-NOMA system. We have considered a generalized CSI error model. Due to BackCom communication, the uncertainties in channels cannot be incorporated in the traditional way as done in the worst-case robust optimization approach. We leveraged Lagrange duality to arrive at a safe and tractable convex reformulation of the robust optimization problem, which is iteratively solved using majorization-minimization framework. In order to tackle the conservative nature of the proposed solution, a slack variables based penalized reformulation was also introduced. Extensive numerical results showcase the superiority of robust H-NOMA with BackCom transmission in the presence of channel imperfections. In particular, the robust version of H-NOMA is not only power efficient but also ensures that the users exceed the minimum data rate requirements most of the time in the uplink.
\balance
\appendices
\section{Proof of Theorem~\ref{thm1:1stApp}}
\label{Proof_thm1}
We focus on the first inequality in \eqref{sinrII} 
\begin{align}
&\min_{\boldsymbol{\gamma}_u\in\mathcal{U}^u,\boldsymbol{\gamma}_{ut}\in\mathcal{U}^{ut}}|h_u|^2|g_{ut}|^2\geq \frac{\epsilon_{ut}^2}{P_{ut}}, 1\leq t\leq u,1\leq u\leq U\label{sinrIIa}\\&\Leftrightarrow
\min_{\boldsymbol{\gamma}_u\in\mathcal{U}^u,\boldsymbol{\gamma}_{ut}\in\mathcal{U}^{ut}}(|\bar{h}_u|^2+\sum_{j=1}^L\gamma_j^ua_j^u)(|\bar{g}_{ut}|^2+\sum_{k=1}^L\gamma_k^{ut}b_k^{ut})\geq\nonumber\\&\hspace{3.5 cm} \frac{\epsilon_{ut}^2}{P_{ut}},1\leq t\leq u,1\leq u\leq U\label{sinrIIb}\\&\Leftrightarrow|\bar{h}_u|^2|\bar{g}_{ut}|^2+\min_{\boldsymbol{\gamma}_u\in\mathcal{U}^u,\boldsymbol{\gamma}_{ut}\in\mathcal{U}^{ut}}(|\bar{h}_u|^2\sum_{k=1}^L\gamma_k^{ut}b_k^{ut}+\nonumber\\&|\bar{g}_{ut}|^2\sum_{j=1}^L\gamma_j^ua_j^u+\sum_{j=1}^L\sum_{k=1}^L\gamma_j^ua_j^u\gamma_k^{ut}b_k^{ut})\geq \frac{\epsilon_{ut}^2}{P_{ut}},1\leq t\leq u,\nonumber\\&\hspace{5cm}1\leq u\leq U\label{sinrIIc}\\&\Leftrightarrow|\bar{h}_u|^2|\bar{g}_{ut}|^2+\min_{\boldsymbol{\gamma}_u\in\mathcal{U}^u,\boldsymbol{\gamma}_{ut}\in\mathcal{U}^{ut}}(\boldsymbol{\gamma}_{ut}^T\mathbf{b}_{ut}+\boldsymbol{\gamma}_{u}^T\mathbf{a}_{u}+\nonumber\\&\hspace{1.0cm}\boldsymbol{\gamma}_{u}^T\mathbf{Q}_{ut}\boldsymbol{\gamma}_{ut})\geq \frac{\epsilon_{ut}^2}{P_{ut}},1\leq t\leq u,1\leq u\leq U\label{sinrIId}
\end{align}
where $\mathbf{b}_{ut}=|\bar{h}_u|^2[b_1^{ut},b_2^{ut},\ldots,b_L^{ut}]^T,\mathbf{a}_{u}=|\bar{g}_{ut}|^2[a_1^{u},a_2^{u},\ldots,a_L^{u}]^T$ and $\mathbf{Q}_{ut}=\boldsymbol{\alpha}_u\boldsymbol{\kappa}_{ut}^T$. We dualize the minimization problem in \eqref{sinrIId} over $\boldsymbol{\gamma}_{u}$.
\begin{align}
&\hspace{0 mm}|\bar{h}_u|^2|\bar{g}_{ut}|^2+\min_{\boldsymbol{\gamma}_u\in\mathcal{U}^u,\boldsymbol{\gamma}_{ut}\in\mathcal{U}^{ut}}(\boldsymbol{\gamma}_{ut}^T\mathbf{b}_{ut}+\boldsymbol{\gamma}_{u}^T(\mathbf{a}_{u}+\mathbf{Q}_{ut}\boldsymbol{\gamma}_{ut}))\geq\nonumber\\&\hspace{3.5cm} \frac{\epsilon_{ut}^2}{P_{ut}},1\leq t\leq u,1\leq u\leq U\label{sinrIIe}\\&\hspace{-2.0mm}\Leftrightarrow|\bar{h}_u|^2|\bar{g}_{ut}|^2+\min_{\boldsymbol{\gamma}_{ut}\in\mathcal{U}^{ut}}\min_{\boldsymbol{\gamma}_u\in\mathcal{U}^u}(\boldsymbol{\gamma}_{ut}^T\mathbf{b}_{ut}+\boldsymbol{\gamma}_{u}^T(\mathbf{a}_{u}+\mathbf{Q}_{ut}\boldsymbol{\gamma}_{ut}))\geq \nonumber\\&\hspace{3.5 cm}\frac{\epsilon_{ut}^2}{P_{ut}},1\leq t\leq u,1\leq u\leq U\label{sinrIIf}\\&\hspace{-2.1mm}\Leftrightarrow |\bar{h}_u|^2|\bar{g}_{ut}|^2+\min_{\boldsymbol{\gamma}_{ut}\in\mathcal{U}^{ut}} \boldsymbol{\gamma}_{ut}^T\mathbf{b}_{ut}+\nonumber\\&\hspace{-2.1mm}\max_{\boldsymbol{\lambda}_u\geq 0,\boldsymbol{\theta}_u\geq 0}\{-\boldsymbol{\lambda}_u^T\mathbf{l}-\boldsymbol{\theta}_u^T\mathbf{u}|\mathbf{a}_u^T+\mathbf{Q}_{ut}\boldsymbol{\gamma}_{ut}-\boldsymbol{\lambda}_u^T\mathbf{A}+\boldsymbol{\theta}_u^T\mathbf{B}= 0\}\geq \nonumber\\&\hspace{3.5 cm}\frac{\epsilon_{ut}^2}{P_{ut}},1\leq t\leq u,1\leq u\leq U\label{sinrIIg}\\&\hspace{-2.1mm}\Leftrightarrow \exists \boldsymbol{\lambda}_u,\boldsymbol{\theta}_u:|\bar{h}_u|^2|\bar{g}_{ut}|^2+\!\!\!\min_{\boldsymbol{\gamma}_{ut}\in\mathcal{U}^{ut}}\!\!\begin{cases}\boldsymbol{\gamma}_{ut}^T\mathbf{b}_{ut}-\boldsymbol{\lambda}_u^T\mathbf{l}-\boldsymbol{\theta}_u^T\mathbf{u}\geq\\\frac{\epsilon_{ut}^2}{P_{ut}},1\leq t\leq u,1\leq u\leq U,\\\mathbf{a}_u^T+\mathbf{Q}_{ut}\boldsymbol{\gamma}_{ut}-\boldsymbol{\lambda}_u^T\mathbf{A}+\\\boldsymbol{\theta}_u^T\mathbf{B}= 0,\boldsymbol{\lambda}_u\geq 0, \boldsymbol{\theta}_u\geq 0,\\1\leq t\leq u,1\leq u\leq U.\end{cases}\label{sinrIIh}\\&\hspace{-2.1mm}\Leftrightarrow \exists \boldsymbol{\lambda}_u:|\bar{h}_u|^2|\bar{g}_{ut}|^2+\min_{\boldsymbol{\gamma}_{ut}\in\mathcal{U}^{ut}}\begin{cases}\boldsymbol{\gamma}_{ut}^T\mathbf{b}_{ut}-\boldsymbol{\lambda}_u^T\mathbf{l}\geq\frac{\epsilon_{ut}^2}{P_{ut}},\\1\leq t\leq u,1\leq u\leq U,\\-(a_u^w+\mathbf{q}_{ut}^w\boldsymbol{\gamma}_{ut})+\boldsymbol{\lambda}_u^T\mathbf{a}^w\geq\\0,\boldsymbol{\lambda}_u\geq 0,1\leq w\leq L,\\1\leq t\leq u,1\leq u\leq U.\end{cases}\label{sinrIIi}
\end{align}
where the equivalence in \eqref{sinrIIi} follows by noting that since $\mathbf{B}$ is invertible, so the equality constraint in \eqref{sinrIIh} can be equivalently written as,
\begin{align}
\boldsymbol{\theta}_u^T= -(\mathbf{a}_u^T+\mathbf{Q}_{ut}\boldsymbol{\gamma}_{ut})\mathbf{B}^{-1}+\boldsymbol{\lambda}_u^T\mathbf{A}\mathbf{B}^{-1}\label{sinrIIi1}
\end{align}
which further due to the non-negativity of $\boldsymbol{\theta}_u$ yields the inequality 
\begin{align}
\boldsymbol{\lambda}_u^T\mathbf{A}\geq(\mathbf{a}_u^T+\mathbf{Q}_{ut}\boldsymbol{\gamma}_{ut}).\label{sinrIIi2}
\end{align}
Now by representing $a_u^w$ as the $w^{\text{th}}$ element of $\mathbf{a}_u$, $\mathbf{q}_{ut}^w$ as the $w^{\text{th}}$ row of $\mathbf{Q}_{ut}$ and $\mathbf{a}^w$ as the $w^{\text{th}}$ column of $\mathbf{A}$, the formulation in \eqref{sinrIIi} follows.
Hence, the first constraint in \eqref{sinrII} is equivalently represented as the following system
\begin{align}
\hspace{-1.1 mm}\exists \boldsymbol{\lambda}_u:|\bar{h}_u|^2|\bar{g}_{ut}|^2+\begin{cases}-\rho\|\mathbf{b}_{ut}\|-\boldsymbol{\lambda}_u^T\mathbf{l}\geq\frac{\epsilon_{ut}^2}{P_{ut}},1\leq t\leq u,\\1\leq u\leq U\\-a_u^w-\rho\|\mathbf{q}_{ut}^w\|_2+\boldsymbol{\lambda}_u^T\mathbf{a}^w\geq 0,1\leq w\leq L,\\\boldsymbol{\lambda}_u\geq 0,1\leq t\leq u,1\leq u\leq U\end{cases}\label{sinrIIeqa_fin}
\end{align}
where we have used the Cauchy-Schwarz inequality to obtain the minimum value over $\boldsymbol{\gamma}_{ut}\in\mathcal{U}^{ut}$.

\bibliographystyle{IEEEtran}
\bibliography{IEEEabrv,Robust_Hybrid_NOMA}

\begin{thebibliography}{10}
\providecommand{\url}[1]{#1}
\csname url@samestyle\endcsname
\providecommand{\newblock}{\relax}
\providecommand{\bibinfo}[2]{#2}
\providecommand{\BIBentrySTDinterwordspacing}{\spaceskip=0pt\relax}
\providecommand{\BIBentryALTinterwordstretchfactor}{4}
\providecommand{\BIBentryALTinterwordspacing}{\spaceskip=\fontdimen2\font plus
\BIBentryALTinterwordstretchfactor\fontdimen3\font minus \fontdimen4\font\relax}
\providecommand{\BIBforeignlanguage}[2]{{%
\expandafter\ifx\csname l@#1\endcsname\relax
\typeout{** WARNING: IEEEtran.bst: No hyphenation pattern has been}%
\typeout{** loaded for the language `#1'. Using the pattern for}%
\typeout{** the default language instead.}%
\else
\language=\csname l@#1\endcsname
\fi
#2}}
\providecommand{\BIBdecl}{\relax}
\BIBdecl

\bibitem{DingNGMA}
Z.~Ding, R.~Schober, P.~Fan, and H.~V. Poor, ``Next generation multiple access for {IMT} towards 2030 and beyond,'' \emph{Sci. China Inf. Sci.}, vol.~67, no.~6, pp. 1--3, 2024.

\bibitem{DingHNoma}
\BIBentryALTinterwordspacing
Z.~Ding, ``Backcom assisted hybrid {NOMA} uplink transmission for ambient {IoT},'' \emph{arXiv}, 2024. [Online]. Available: \url{https://arxiv.org/abs/2403.16498}
\BIBentrySTDinterwordspacing

\bibitem{DingMECHNoma}
Z.~Ding, D.~Xu, R.~Schober, and H.~V. Poor, ``Hybrid {NOMA} offloading in multi-user {MEC} networks,'' \emph{IEEE Trans. Wireless Commun.}, vol.~21, no.~7, pp. 5377--5391, 2022.

\bibitem{Ding_MEC}
Z.~{Ding}, J.~{Xu}, O.~A. {Dobre}, and H.~V. {Poor}, ``Joint power and time allocation for {NOMA-MEC} offloading,'' \emph{IEEE Trans. Veh. Technol.}, vol.~68, no.~6, pp. 6207--6211, 2019.

\bibitem{Ding-MEC1}
Z.~Ding, D.~W.~K. Ng, R.~Schober, and H.~V. Poor, ``Delay minimization for {NOMA-MEC} offloading,'' \emph{IEEE Signal Process. Lett.}, vol.~25, no.~12, pp. 1875--1879, 2018.

\bibitem{Liu-HNoma}
L.~Liu, B.~Sun, Y.~Wu, and D.~H.~K. Tsang, ``Latency optimization for computation offloading with hybrid {NOMA-OMA} transmission,'' \emph{IEEE Internet Things J.}, vol.~8, no.~8, pp. 6677--6691, 2021.

\bibitem{DingDLNoma}
\BIBentryALTinterwordspacing
Z.~Ding, R.~Schober, and H.~V. Poor, ``Design of downlink hybrid {NOMA} transmission,'' \emph{arXiv}, 2024. [Online]. Available: \url{https://arxiv.org/abs/2401.16965}
\BIBentrySTDinterwordspacing

\bibitem{Cirine}
C.~Chaieb, F.~Abdelkefi, and W.~Ajib, ``Deep reinforcement learning for resource allocation in multi-band and hybrid {OMA-NOMA} wireless networks,'' \emph{IEEE Trans. Commun.}, vol.~71, no.~1, pp. 187--198, 2023.

\bibitem{Lei}
J.~Lei, T.~Zhang, and Y.~Liu, ``Hybrid {NOMA} for {STAR-RIS} enhanced communication,'' \emph{IEEE Trans. Veh. Technol.}, vol.~73, no.~1, pp. 1497--1502, 2024.

\bibitem{WangAoI}
Q.~Wang, H.~Chen, Y.~Li, and B.~Vucetic, ``Minimizing age of information via hybrid {NOMA/OMA},'' in \emph{Proc. of IEEE International Symposium on Information Theory (ISIT)}, 2020, pp. 1753--1758.

\bibitem{Lyu}
W.~Lyu, Y.~Xiu, X.~Li, S.~Yang, P.~L. Yeoh, Y.~Li, and Z.~Zhang, ``Hybrid {NOMA} assisted integrated sensing and communication via {RIS},'' \emph{IEEE Trans. Veh. Technol}, vol.~73, no.~5, pp. 7368--7373, 2024.

\bibitem{ZhuURLLC}
Y.~Zhu, X.~Yuan, Y.~Hu, T.~Wang, M.~C. Gursoy, and A.~Schmeink, ``Low-latency hybrid {NOMA-TDMA}: {QoS}-driven design framework,'' \emph{IEEE Trans. Wireless Commun.}, vol.~22, no.~5, pp. 3006--3021, 2023.

\bibitem{Sultana}
T.~Sultana and S.~Dumitrescu, ``Globally optimal max-min rate joint channel and power allocation for hybrid {NOMA-OMA} downlink systems,'' \emph{IEEE Trans. Signal Process}, pp. 1--16, 2025, {E}arly Access.

\bibitem{Liebhart}
R.~Liebhart, M.~Shafi, H.~Tataria, G.~Shivanandan, and D.~handramouli, ``Perspectives on {6G} architectures,'' \emph{IEEE Wireless Commun.}, vol.~32, no.~1, pp. 108--114, 2025.

\bibitem{Ping}
J.-P. Niu and Y.~G. Li, ``An overview on backscatter communications,'' \emph{Journal of Communications and Information Networks}, vol.~4, no.~2, pp. 1--14, 2019.

\bibitem{AharonNemirovskiGhaoui2009}
A.~Ben-Tal, L.~E. Ghaoui, and A.~Nemirovski, \emph{Robust Optimization}.\hskip 1em plus 0.5em minus 0.4em\relax NJ, Princeton: Princeton Univ. Press, 2009.

\bibitem{Biguesh}
M.~Biguesh, S.~Shahbazpanahi, and A.~B. Gershman, ``Robust downlink power control in wireless cellular systems,'' \emph{EURASIP J. Wireless Comm. Netw.}, no.~2, pp. 261--272, 2004.

\bibitem{FangImp}
F.~Fang, K.~Wang, Z.~Ding, and V.~C.~M. Leung, ``Energy-efficient resource allocation for {NOMA-MEC} networks with imperfect {CSI},'' \emph{IEEE Trans. Commun.}, vol.~69, no.~5, pp. 3436--3439, 2021.

\bibitem{Fang}
------, ``Energy-efficient resource allocation for noma-mec networks with imperfect {CSI},'' \emph{IEEE Trans. Commun.}, vol.~69, no.~5, pp. 3436--3449, 2021.

\bibitem{Marton}
M.~Nasz\'{o}di, F.~Nazarov, and D.~Ryabogin, ``Fine approximation of convex bodies by polytopes,'' \emph{American Journal of Mathematics}, vol. 142, no.~3, pp. 809--820, 2020.

\bibitem{Bronshteyn}
E.~M. Bronshteyn and L.~D. Ivanov, ``The approximation of convex sets by polyhedra,'' \emph{Siberian Mathematical Journal}, vol.~16, pp. 852--853, 1975.

\bibitem{JeffAnd_Imp_SIC}
J.~G. Andrews and T.~H. Meng, ``Optimum power control for successive interference cancellation with imperfect channel estimation,'' \emph{IEEE Trans. Wireless Commun.}, vol.~2, no.~2, pp. 375--383, 2003.

\bibitem{HanifTran_Robust}
M.~F. Hanif, L.-N. Tran, M.~Juntti, and S.~Glisic, ``On linear precoding strategies for secrecy rate maximization in multiuser multiantenna wireless networks,'' \emph{IEEE Trans. Signal Process.}, vol.~62, no.~14, pp. 3536--3551, 2014.

\bibitem{BoydCVX2004}
S.~Boyd and L.~Vandenberghe, \emph{Convex Optimization}.\hskip 1em plus 0.5em minus 0.4em\relax Cambridge, U.K.: Cambridge Univ. Press, 2004.

\bibitem{PalomarMM}
Y.~Sun, P.~Babu, and D.~P. Palomar, ``Majorization-minimization algorithms in signal processing, communications, and machine learning,'' \emph{IEEE Trans. Signal Process.}, vol.~65, no.~3, pp. 794--816, 2017.

\bibitem{Inosha_Topology}
I.~Sugathapala, M.~F. Hanif, B.~Lorenzo, S.~Glisic, M.~Juntti, and L.-N. Tran, ``Topology adaptive sum rate maximization in the downlink of dynamic wireless networks,'' \emph{IEEE Trans. Commun.}, vol.~66, no.~8, pp. 3501--3516, 2018.

\bibitem{BoydDiffCvx}
T.~Lipp and S.~Boyd, ``Variations and extension of the convex-concave procedure,'' \emph{Optim. Eng.}, vol.~17, no.~2, pp. 263--287, 2016.

\bibitem{HunterMM}
D.~R. Hunter and K.~Lange, ``A tutorial on {MM} algorithms,'' \emph{Amer. Statist.}, vol.~58, no.~1, pp. 30--37, 2004.

\bibitem{MarksInnerApprox}
B.~R. Marks and G.~P. Wright, ``A general inner approximation algorithm for nonconvex mathematical programs,'' \emph{Oper. Res.}, vol.~26, no.~4, pp. 681--683, 1978.

\bibitem{Thi_1}
H.~A.~L. Thi, V.~N. Huynh, and T.~P. Dinh, ``{DC} programming and {DCA} for general {DC} programs,'' in \emph{Adv. Comput, Methods Knowl. Eng.}, 2014, pp. 15--35.

\bibitem{KumarNOMA}
V.~Kumar, M.~F. Hanif, M.~Juntti, and L.-N. Tran, ``A max-min task offloading algorithm for mobile edge computing using non-orthogonal multiple access,'' \emph{IEEE Trans. Veh. Technol.}, vol.~72, no.~9, pp. 12\,332--12\,337, 2023.

\bibitem{Thi_2}
H.~A.~L. Thi and T.~P. Dinh, ``{DC} programming and {DCA}: thirty years of developments,'' \emph{Math. Program.}, vol. 169, no.~1, pp. 5--68, 2018.

\bibitem{AharonNemirovski2001}
A.~Ben-Tal and A.~Nemirovski, \emph{Lectures on Modern Convex Optimization}.\hskip 1em plus 0.5em minus 0.4em\relax Society for Industrial and Applied Mathematics, 2001.

\bibitem{mosek1}
\BIBentryALTinterwordspacing
{MOSEK ApS}, \emph{The {MOSEK} {API} for {MATLAB} manual. Version 11.0.16}, 2025. [Online]. Available: \url{https://docs.mosek.com/latest/matlabapi/index.html}
\BIBentrySTDinterwordspacing

\bibitem{cvx}
M.~Grant and S.~Boyd, ``{CVX}: Matlab software for disciplined convex programming, version 2.2,'' \url{http://cvxr.com/cvx}, Apr. 2024.

\end{thebibliography}
\end{document}